\documentclass[aps,prl,amsmath,amssymb,twocolumn,showpacs]{revtex4-1}
\usepackage[english]{babel}
\usepackage{graphicx}
\usepackage[dvipdfm]{hyperref}
\usepackage{hyperref}

\newcommand{\abbrev}[1]{\textsc{#1}}

\begin{document}

\title{Phase diagram of quasi-two-dimensional bosons in laser speckle potential}

\author{G.~E.~Astrakharchik$^1$}
\author{K.~V.~Krutitsky$^2$}
\author{P.~Navez$^{2,3}$}
\affiliation{$^1$ Departament de F\'{\i}sica i Enginyeria Nuclear, Campus Nord B4-B5,
                  Universitat Polit\`ecnica de Catalunya, E-08034 Barcelona, Spain}
\affiliation{$^2$ Fakult\"at f\"ur Physik der Universit\"at Duisburg-Essen, Campus Duisburg,
                  Lotharstrasse 1, 47048 Duisburg, Germany}
\affiliation{$^3$ Institut f\"ur Theoretische Physik, TU Dresden, 01062 Dresden, Germany}

\date{10 June 2013}

\begin{abstract}
We have studied the phase diagram of a quasi-two-dimensional interacting Bose gas at zero temperature in the presence of random potential created by laser speckles. The superfluid fraction and the fraction of particles with zero momentum are obtained within the mean-field Gross-Pitaevskii theory and in diffusion Monte Carlo simulations. We find a transition from the superfluid to the insulating state, when the strength of the disorder grows. Estimations of the critical parameters are compared with the predictions of the percolation theory in the Thomas-Fermi approximation. Analytical expressions for the zero-momentum fraction and the superfluid fraction are derived in the limit of weak disorder and weak interactions within the framework of the Bogoliubov theory. Limits of validity of various approximations are discussed.
\end{abstract}


\maketitle

\emph{Introduction.---}
Ultracold atoms provide an irreplaceable tool to study interacting quantum systems with disorder
as they offer a unique opportunity to control the system~\cite{ReviewDisorder}. Interaction strength can be tuned by Feshbach resonances, while disorder with known statistical properties and tunable parameters can be created either by optical means using incommensurate optical lattices and laser speckles~\cite{DZSZL03} or through the interaction with impurities localized at random positions~\cite{GPRVS11}. By using confining potentials in different spatial directions, one can also control the system dimensionality.

Laser speckles allow the creation of unbounded random potentials with a finite correlation length~\cite{G,CVRSAP06}. In recent years, laser speckles were used for the experimental observation of Anderson localization of bosons in one dimension (1D)~\cite{BJZBHLCSBA} and three dimensions (3D)~\cite{JBMCJPPSAB} as well as of fermions in 3D~\cite{KMZD11}, suppression of transport in elongated 1D  geometry~\cite{LFMWFI05,CVHRBSGSA05,FFGLMWI05,CVRSAP06} and Bose-Einstein condensate in 3D optical lattices~\cite{WPMZCD09}. Coherent phenomena and diffusion of cold atoms as well as superfluidity of Feshbach molecules were also experimentally studied in quasi-two-dimensional (quasi-2D) geometry~\cite{RBAPPSABB10,BRHR11,APHSABB11,JMRDPBAJ12,KSMBE}.

The properties of single-particle eigenstates in the presence of laser speckles are almost the same as in the case of uncorrelated disorder. The difference was found only in 1D, where there is a crossover from exponential localization to the algebraic one~\cite{BJZBHLCSBA,SCLBSA}, while all the eigenstates in the presence of uncorrelated disorder are exponentially localized. In 2D, the wave functions always show an exponential long-range decay and one has to distinguish between weak and strong localization~\cite{KMDSM,KSMDM,MKDM}. In 3D, there is a critical energy that separates extended and localized states (mobility edge) but its precise determination still remains a difficult problem~\cite{PPS,PPSP}.

The quantum many-body problem of interacting particles in the presence of disorder is even more challenging. Most of the studies of the interacting bosons in the presence of laser speckles are done within the framework of the mean-field theory. The Gross-Pitaevskii equation (GPE) was used to study ground-state density profiles in 1D~\cite{S06}, expansion in 1D~\cite{M06,SCLBSA,PS08,SCLBA08,PLBAS11}, transport of quasi-1D Bose-Einstein condensate (BEC) in the subsonic and supersonic regimes~\cite{PASLP09}, and dipole and quadrupole collective oscillations in 3D~\cite{M06}. The Thomas-Fermi approximation was successfully employed in the regime when the correlation length of the disorder potential is much larger than the healing length of the condensate~\cite{S06,CVRSAP06,SCLBA08,PGMP10,B12}. Combined with the percolation theory~\cite{PGMP10,PRBBAPAPS11}, this approximation was used to estimate the critical values of parameters for the superfluid-insulator transition in a  3D system~\cite{PGMP10}.

Quantum fluctuations were taken into account within the framework of the Bogoliubov theory that predicts Anderson localization of quasiparticles~\cite{LCBAS07,LS11}. The Bogoliubov theory was also used to study excitation spectrum, corrections to the speed of sound and condensate population in the case of weak disorder~\cite{GM11,MG12,GM12p}. GPE and Bogoliubov-de~Gennes equations in 1D reveal the superfluid--Bose-glass transition~\cite{FWS09,FWS10,LCBALS07}. Quantum Monte Carlo (QMC) calculations in 3D were performed in Refs.~\cite{PGP09,PGMP10}, where the finite temperature superfluid-insulator transition was studied. Note that the existence of the superfluid-insulator transition in the continuum models is in agreement with the predictions of the disordered Bose-Hubbard model~\cite{FWGF89,PS98,RSZ99,GPPST09,SKPS11}.

\begin{figure}[t]
\includegraphics[angle=0,width=0.8\columnwidth]{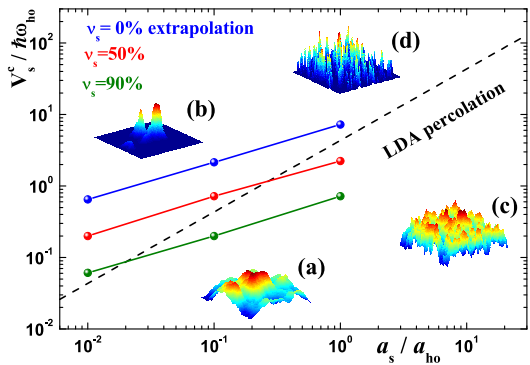}
\caption{(Color online)
         Phase diagram.
         Solid upper (blue) line is the boundary between the superfluid ($\nu_{\rm s}>0$) and insulator ($\nu_{\rm s}=0$) phases.
         The solid middle (red) and bottom (green) lines are calculated from the requirement that the superfluid fraction
         takes fixed values $\nu_{\rm s}=0.5$ and $\nu_{\rm s}=0.9$, respectively.
         The dashed line is the estimate of the critical values of $V_{\rm s}$
         from the percolation model in the local-density approximation
         [Eq.~(\ref{crit-perc})].
         The insets show density profiles calculated for one disorder realization with
         (a)~$V_{\rm s}/\hbar\omega_{\rm ho}=0.1$, $a_{\rm s}/a_{\rm ho}=0.01$;
         (b)~$V_{\rm s}/\hbar\omega_{\rm ho}=1$,   $a_{\rm s}/a_{\rm ho}=0.01$;
         (c)~$V_{\rm s}/\hbar\omega_{\rm ho}=1$,   $a_{\rm s}/a_{\rm ho}=1$; and
         (d)~$V_{\rm s}/\hbar\omega_{\rm ho}=40$,  $a_{\rm s}/a_{\rm ho}=1$.
        }
\label{pd}
\end{figure}

In the present work, we investigate the superfluidity and the off-diagonal long-range order of interacting bosons at zero temperature in the presence of laser speckles in a quasi-2D continuum setup. On the basis of QMC calculations we find a transition from the superfluid to the insulator phase. The dependence of the critical disorder strength $V_{\rm s}^{\rm c}$ on the interaction strength $a_{\rm s}$ is shown in Fig.~\ref{pd} by upper (solid blue) line. The lines that determine the dependence of the disorder strength $V_{\rm s}$ on the interaction strength $a_{\rm s}$ at fixed values of the superfluid fraction $\nu_s$ are also displayed in Fig.~\ref{pd}.

\emph{The system.---}
The Hamiltonian of $N$ interacting bosons in the presence of an external random potential $V({\bf r})$ and a tight harmonic trapping in the $z$ direction is
\begin{equation}
H
=
\sum_{i=1}^N
\left[
    - \frac{\hbar^2\nabla_i^2}{2m}
    +
    V({\bf r}_i)
    +
    \frac{m}{2}
    \omega_{\rm ho}^2 z_i^2
\right]
+
\sum_{i<j}
V_{\rm pp}(\mathbf{r}_i-\mathbf{r}_j)
\;.
\nonumber
\end{equation}
We assume that the frequency $\omega_{\rm ho}$ is so large, that particles do not populate higher levels of the trap. Under this realistic assumption, the system is effectively quasi-two-dimensional. We consider a system of a finite size $L\times L$ in the $(x,y)$-plane. The interaction potential $V_{\rm pp}(\mathbf{r})$ is chosen to be of a short-range type as a relevant model for ultracold atoms. In the QMC simulations, we model those as hard spheres of the diameter $a_{\rm s}$. In the dilute regime, the potential can be replaced by a contact term $4\pi\hbar^2 a_{\rm s}/m~\delta(\mathbf{r})$ that is used in the GPE. For both interaction potentials, $a_{\rm s}$ is the $s$-wave scattering length.

The external random potential $V({\bf r})$ is assumed to be created by laser speckles~\cite{G,CVRSAP06}. In the presence of a strong harmonic confinement one can neglect its $z$ dependence. We numerically generate two-dimensional speckle patterns $V({\bf r})\equiv V({\bf r}_\perp)$ following the procedure described in Ref.~\cite{H}. Each realization of $V({\bf r}_\perp)$ is a periodic function of $x$ and $y$, therefore, permits us to impose periodic boundary conditions in the $(x,y)$ plane. The random potential $V({\bf r}_\perp)$ has exponential probability distribution with the mean value $V_{\rm s}=\overline{V({\bf r}_\perp)}$ which is an important parameter of the problem. The explicit form of the autocorrelation function
\begin{equation}
\label{af}
f({\bf r}_\perp-{\bf r}_\perp')
=
\overline{ V({\bf r}_\perp) V({\bf r}_\perp') }
-
\overline{V({\bf r}_\perp)}^2
\end{equation}
is determined by the aperture of the laser system~\cite{G}. In the case of a square aperture and in the limit $L\to\infty$, it is given by
\begin{equation}
\label{corrsquare}
f({\bf r}_\perp-{\bf r}_\perp')
=
V_{\rm s}^2
{\rm sinc}^2
\left(
    \frac{x-x'}{L_{\rm c}}
\right)
\,
{\rm sinc}^2
\left(
    \frac{y-y'}{L_{\rm c}}
\right)
\;,
\end{equation}
where ${\rm sinc}(x)=(\sin \pi x)/(\pi x)$ and $L_{\rm c}$ is the correlation length that determines the characteristic size of the speckles. In the case of a circular aperture, it takes the form
\begin{equation}
\label{corrcircle}
f({\bf r}_\perp-{\bf r}_\perp')
=
V_{\rm s}^2
\left[
    2
    \frac
    {J_1(x_0 |{\bf r}_\perp-{\bf r}_\perp'|/L_{\rm c})}
    {x_0 |{\bf r}_\perp-{\bf r}_\perp'|/L_{\rm c}}
\right]^2
\;,
\end{equation}
where $x_0\approx 3.83171$ is the first nontrivial zero of the Bessel function $J_1(x)$.

\emph{Numerical methods.---}
For the numerical solution of the problem we employ variational and diffusion QMC methods~\cite{MC}. In the variational QMC calculations, the trial many-body wave function is constructed in the form
\begin{equation}
\Psi_{\rm T}(\mathbf{r}_1,\dots,\mathbf{r}_N)
=
\prod_{i=1}^N \Phi(\mathbf{r}_i)
\prod_{i<j} f_2(|{\bf r}_i-{\bf r}_j|)
\;,
\nonumber
\end{equation}
where $f_2(|{\bf r}|)$ is a two-body Jastrow term. The strong harmonic confinement in the $z$ direction allows us to represent the one-body term as $\Phi({\bf r})=\psi({\bf r}_\perp)\psi_\text{ho}(z)$, where $\psi_\text{ho}(z)$ is the ground-state wave function of the harmonic oscillator and $\psi({\bf r}_\perp)$ satisfies the two-dimensional GPE
\begin{equation}
\label{GPE}
\mu\psi({\bf r}_\perp)
=
\left[
    -\frac{\hbar^2 \nabla_\perp^2}{2m}
    +
    V({\bf r}_\perp)
    +
    g_2 N
    \left|
        \psi({\bf r}_\perp)
    \right|^2
\right]
\psi({\bf r}_\perp)
\;
\end{equation}
with the effective coupling constant $g_2=\sqrt{8\pi}\hbar^2 a_{\rm s}/(m a_{\rm ho})$. Equation~(\ref{GPE}) was solved numerically on a grid with the spatial step less than $L_c/10$ using the imaginary-time propagation technique. The values of $\psi({\bf r}_\perp)$ at intermediate points needed for the QMC sampling were obtained by a polynomial interpolation.

The superfluid fraction $\nu_{\rm s}$ is obtained in QMC simulations using the winding-number technique~\cite{PC87}. It can be also obtained from the GPE in a moving reference frame introduced by means of substitution $\psi({\bf r}_\perp)=\phi({\bf r}_\perp)\exp(i{\bf k}_{\rm s}\cdot{\bf r}_\perp)$ in Eq.~(\ref{GPE}), where $\phi({\bf r}_\perp)$ satisfies periodic boundary conditions. For small ${\bf k}_{\rm s}$, the expectation value of the momentum ${\bf P}$ has a linear dependence on ${\bf k}_{\rm s}$, i.e., ${\bf P}=N \nu_{\rm s} \hbar{\bf k}_{\rm s}$ which determines the superfluid fraction $\nu_{\rm s}$.

The fraction of particles with zero momentum $\nu_0$, which gives a lower estimate of the condensate fraction~\cite{AK11}, is calculated from the one-body density matrix (OBDM) $\rho_1({\bf r},{\bf r}')$, integrated over $z'=z$, as the asymptotic value at large distance $|{\bf r}_\perp-{\bf r}_\perp'|\to\infty$. In QMC calculations, OBDM is obtained by extrapolation from the variational and mixed estimators, while in the mean-field theory, it is readily obtained from the solution of the GPE as $\rho_1({\bf r},{\bf r}')=\Phi({\bf r})\Phi({\bf r}')$.

\emph{Numerical results.---}
In the problem we consider, there are three natural spatial scales: healing length $\xi=\hbar/\sqrt{mg_2n_2}$, correlation length of the disorder $L_{\rm c}$, and the oscillator length $a_{\rm ho}=\sqrt{\hbar/\left(m\omega_{\rm ho}\right)}$. We use $a_{\rm ho}$ as a length unit and the energies are measured in the units of $\hbar\omega_{\rm ho}$. Our numerical calculations are done for the square aperture with realistic values of $n_2=N/L^2=0.1/a_{\rm ho}^2$ and $L_{\rm c}=a_{\rm ho}$.

The density profiles calculated for one disorder realization in different regimes are shown in Fig.~\ref{pd}. In the case of weak interactions [insets (a) and (b)], corresponding to $\xi\approx 14\,a_{\rm ho}$ larger than $L_{\rm c}$, a rather weak disorder is sufficient to localize the atoms in the region where the random potential has an extended minimum. In the case of strong interactions [insets (c) and (d)], where $\xi\approx 1.4\,a_{\rm ho}$  is comparable to $L_{\rm c}$, the density profile has a multipeak structure with  maxima corresponding to local minima of the disorder potential. We find that the density profiles are very well described by the GPE even for very strong disorder similarly to the 3D case~\cite{PGMP10}.

\begin{figure}[t]
\includegraphics[angle=0,width=0.8\columnwidth]{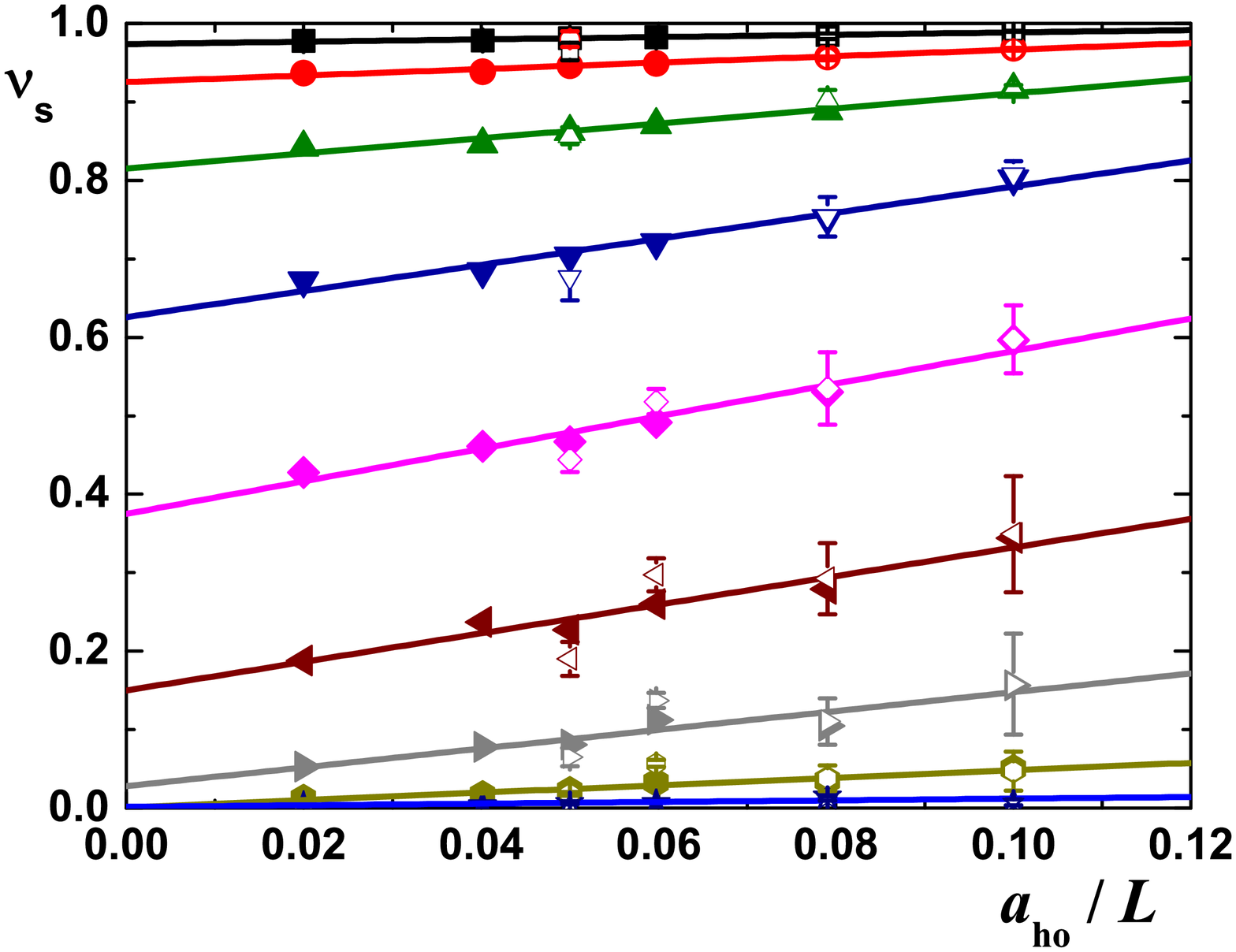}
\caption{(Color online)
         Dependence of the superfluid fraction on the system size for $a_{\rm s}/a_{\rm ho}=0.1$.
         Solid symbols indicate the results obtained from the GPE; open symbols with DMC;
         and straight lines are the fits~(\ref{fit}) corresponding to (from top to the bottom)
         $V_{\rm s}/\hbar\omega_{\rm ho}=0.1$,~$0.18$,~$0.32$,~$0.56$,~$1$,~$1.8$,~$3.2$,~$5.6$,~$10$.
        }
\label{sfL}
\end{figure}

\begin{figure}[t]
\includegraphics[angle=0,width=0.8\columnwidth]{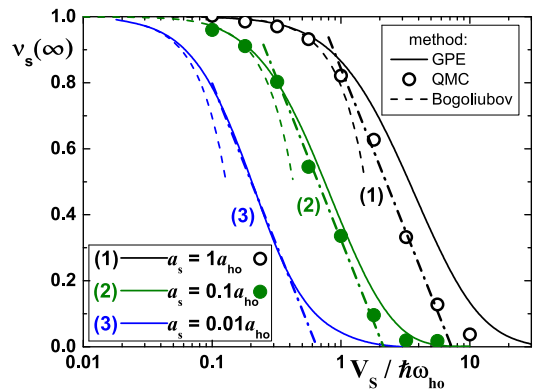}
\caption{(Color online)
         Superfluid fraction in the thermodynamic limit.
         The results obtained from the GPE for $a_{\rm s} / a_{\rm ho} = 1$~(1), $0.1$~(2), $0.01$~(3) are shown by solid lines,
         and the corresponding predictions of the Bogoliubov theory described by Eq.~(\ref{sfsquare})
         are represented by dashed lines labeled by the same numbers.
         QMC data for $a_{\rm s} / a_{\rm ho} = 0.1$ are indicated by filled circles
         and for $a_{\rm s} / a_{\rm ho} = 1$ by open circles.
         For $a_{\rm s} / a_{\rm ho} = 0.01$ we find no difference between GPE and QMC.
         Dash-dotted lines show logarithmic fits used to determine the transition point $V_{\rm s}^{\rm c}$.
         }
\label{sf-inf}
\end{figure}

The superfluid fraction calculated in finite-size systems for $a_{\rm s}/a_{\rm ho}=0.1$ and different values of $V_{\rm s}$ after averaging over several disorder realizations is shown in Fig.~\ref{sfL}. It decreases with the increase of $V_{\rm s}$ and $L$. The size dependence (see Fig.~\ref{sfL}) is well approximated by the fit
\begin{equation}
\label{fit}
\nu_{\rm s}(L)
=
\nu_{\rm s}(\infty) + c/L
\end{equation}
where $\nu_{\rm s}(\infty)$ is the value of the superfluid fraction in the thermodynamic limit and $c$ describes finite-size correction. The dependence of $\nu_{\rm s}(\infty)$ on $V_{\rm s}$ is reported in Fig.~\ref{sf-inf}. The predictions of the GPE are in good quantitative agreement with the QMC data as long as the interactions are not too strong. Quantum fluctuations fully taken into account in the QMC calculations and completely neglected in the GPE become noticeable for large interactions and lead to additional suppression of the superfluidity. Our calculations show that for a fixed strength of disorder $V_s$, stronger interactions lead to a larger superfluid fraction. For a fixed interaction strength, we observe a continuous decrease of $\nu_{\rm s}$ to zero when the disorder strength grows indicating that there is a transition from the superfluid to the insulating phase. The critical strength of the speckle potential $V_{\rm s}\equiv V_{\rm s}^{\rm c}(a_{\rm s})$ is obtained by extrapolating $\nu_{\rm s}$ to zero using logarithmic fit (see  Fig.~\ref{sf-inf}). This procedure leads the phase diagram shown in Fig.~\ref{pd} by the solid blue line, which constitutes the main result of our paper.

\begin{figure}[t]
\includegraphics[angle=0,width=0.8\columnwidth]{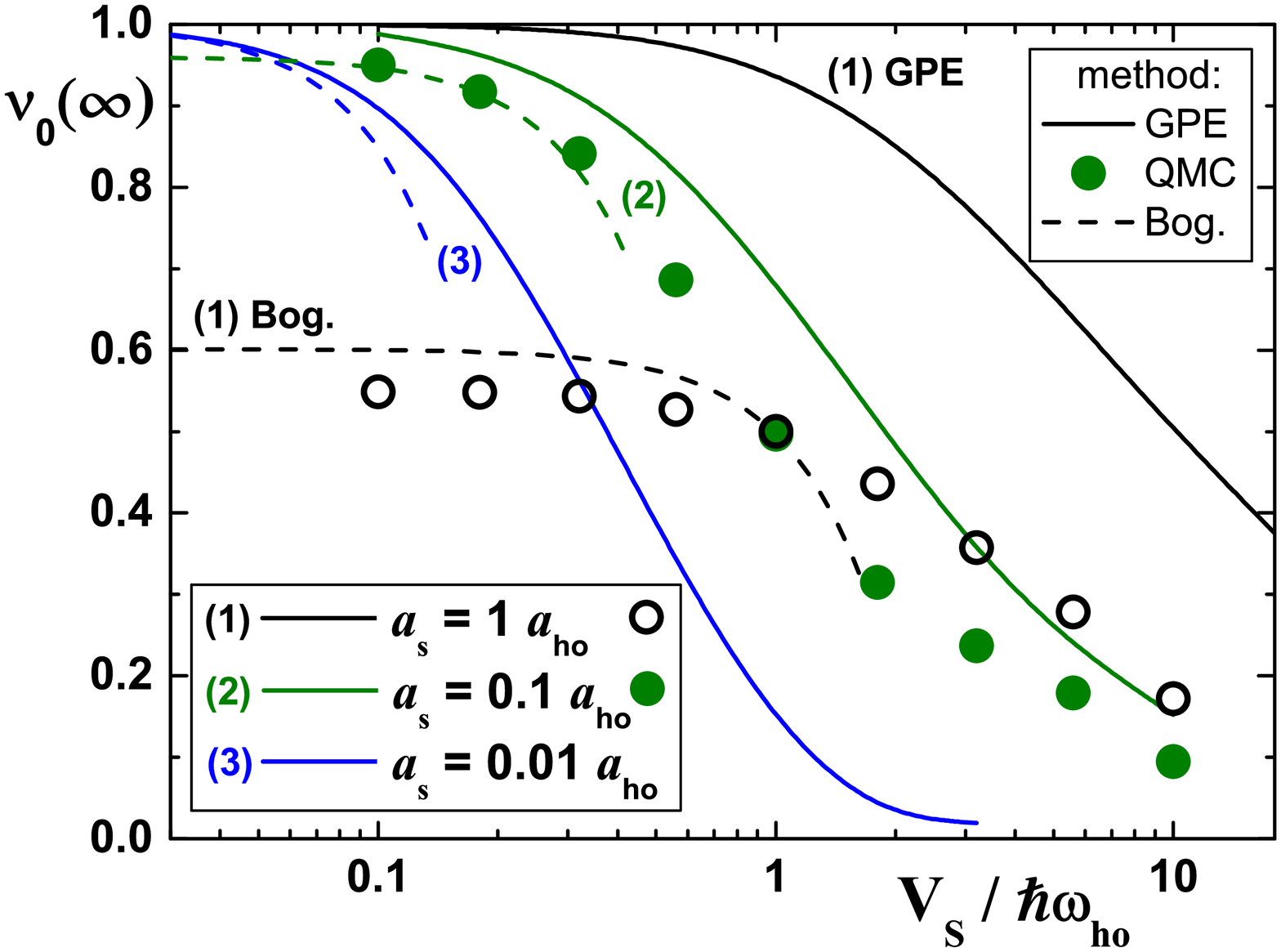}
\caption{(Color online)
         Fraction of particles with zero momentum in the thermodynamic limit.
         The results obtained from the GPE for $a_{\rm s} / a_{\rm ho} = 1$~(1), $0.1$~(2), $0.01$~(3) are shown by solid lines,
         and the corresponding predictions of the Bogoliubov theory are represented by dashed lines.
         QMC data for $a_{\rm s} / a_{\rm ho} = 0.1$ are indicated by filled circles
         and for $a_{\rm s} / a_{\rm ho} = 1$ by open circles.
         For $a_{\rm s} / a_{\rm ho} = 0.01$ we find no difference between GPE and QMC.
        }

\label{k0-inf}
\end{figure}

The fraction of particles with zero momentum shows almost no size dependence for sufficiently large $L$. The thermodynamic values $\nu_0(\infty)$ presented in Fig.~\ref{k0-inf} are obtained by fitting the GPE data for $N\ge 40$ by a constant. With the increase of disorder, $\nu_0$ smoothly decreases similarly to $\nu_{\rm s}$. For stronger interactions, the GPE predicts an increase of $\nu_0$, while exact QMC calculations show in general a non-monotonic dependence of $\nu_0$ on $a_{\rm s}$. The comparison of the results obtained from the GPE and QMC shows that quantum fluctuations play an important role for $a_{\rm s}\gtrsim 0.5\,a_{\rm ho}$ and lead to the suppression of $\nu_0$.

\emph{Bogoliubov theory.---}
For weak external potential $V({\bf r})$, the GPE can be solved analytically in the lowest orders of $V$~\cite{S06,LS11,GM11,AK11,MG12,GM12p}. Assuming that $V({\bf r})$ has a period $L$ in $d$ spatial dimensions and possesses cubic or spherical symmetry, perturbative solution of the GPE in the moving reference frame in the limit of vanishing ${\bf k}_{\rm s}$ leads to the following result for the superfluid fraction:
\begin{equation}
\label{nus}
\nu_{\rm s}
=
1 -
\frac{4}{d}
\sum_{\bf n}
\frac
{
 \left(
     1 - \delta_{{\bf n},{\bf 0}}
 \right)
 \left|\tilde V({\bf n})\right|^2
}
{
 L^d\left[\frac{\hbar^2}{2m}
 \left(\frac{2\pi}{L}\right)^2 {\bf n}^2 + 2 g_d n_d\right]^2
}
\;,
\end{equation}
where $g_d$ is an effective interaction parameter in $d$ dimensions, $n_d=N/L^d$,
\begin{eqnarray}
\label{FT}
\tilde V({\bf n})
=
\frac{1}{L^{d/2}}
\int\limits_{-L/2}^{L/2}
dr_1
\dots
\int\limits_{-L/2}^{L/2}
dr_d
\;
e^{- i \frac{2\pi}{L} {\bf n} \cdot {\bf r}}
V({\bf r})
\;.
\end{eqnarray}
Within the same formalism we also get $\nu_0=1-\frac{d}{4}(1-\nu_s)$. Standard Bogoliubov theory~\cite{disorder} gives exactly the same result for $\nu_{\rm s}$ but $\nu_0$ contains also an additional depletion $-g
_2 m/(4\pi\hbar^2)$~\cite{MG12} due to quantum fluctuations~\cite{LHY} which are neglected in the GPE.

Since the speckle potential is random, we have to perform statistical averaging in Eq.~(\ref{nus}). In general, $\overline{\left|\tilde V({\bf n})\right|^2}$ coincides with the Fourier transform~(\ref{FT}) of the autocorrelation function~(\ref{af}) of the disorder potential. In the case of $\delta$-correlated disorder, this allows us to reproduce the results of Ref.~\cite{disorder}. For laser speckles with the square aperture and in the limit $L\to\infty$, we get
\begin{eqnarray}
\label{sfsquare}
&&
\nu_{\rm s}
=
1
-
\left(
    \frac{V_{\rm s} b}{g_2 n_2}
\right)^2
\left\{
    2
    \sqrt{1+b^2}
    \,
    {\rm arccot}\,\sqrt{1+b^2}
\right.
\\
&&
\left.
    -
    2 b
    \,
    {\rm arccot}\, b
    -
    b^2
    \left[
        \ln b
        -
        \ln
        \left(
            1
            +
            b^2
        \right)
        +
        \frac{1}{2}
        \ln
        \left(
            2
            +
            b^2
        \right)
    \right]
\right\}
\;,
\nonumber
\end{eqnarray}
where $b=L_{\rm c}/(\pi\xi)$. For the circular aperture, we have
\begin{equation}
\label{sfcircle}
\nu_{\rm s}
=
1-
2
\left(
    \frac{V_{\rm s} b}{g_2 n_2}
\right)^2
\left(
    2 b^2
    -
    2 b
    \sqrt{1+b^2}
    +
    1
\right)
\;,
\end{equation}
where $b=L_{\rm c}/(x_0\xi)$. In the limit $\xi\gg L_{\rm c}$, $\nu
_{\rm s}$ tends to unity, while in the opposite limit it takes the asymptotic value
$
\nu_{\rm s}
=
1
-
\frac{1}{2}
\left(
    {V_{\rm s}}/{g_2 n_2}
\right)^2
$.

The predictions of the Bogoliubov theory for $\nu_{\rm s}$ and $\nu_0$ are shown by dashed lines in Figs.~\ref{sf-inf} and~\ref{k0-inf},  respectively. As expected, Bogoliubov theory works better for stronger interactions, although it has a tendency to underestimate $\nu_{\rm s}$ as well as $\nu_0$ already at small disorder strengths. Higher order calculation would probably improve the situation.

\emph{Thomas-Fermi regime and percolation analysis.---}
If $\xi\ll L_{\rm c}$, the density profile of the system in an external potential can be calculated within the Thomas-Fermi approximation and has the following form
\begin{eqnarray}
\label{rhoTF}
\left|
    \psi({\bf r}_\perp)
\right|^2
&=&
\frac
{\mu - V({\bf r}_\perp)}
{g_2 N}
\Theta
\left[
    \mu - V({\bf r}_\perp)
\right]
\;,
\\
\label{muTF}
\mu
&=&
g_2 n_2
+
V_{\rm s}
-
V_{\rm s} \exp(-\mu/V_{\rm s})
\;.
\end{eqnarray}
Equation~(\ref{rhoTF}) allows the existence of the regions where the density completely vanishes. The fraction of space occupied by the particles is $\Omega=1-\exp(-\mu/V_{\rm s})$. If $\Omega$ exceeds the critical value of the percolation transition $\Omega_{\rm c}$, there are infinitely extended regions of space occupied by the particles and the system is superfluid. Otherwise, it is insulating. The percolation analysis can be used for the estimation of the critical values of parameters as was done in 3D~\cite{PGMP10}. Using Eq.~(\ref{muTF}) we get an estimate for the critical value of the  disorder strength
\begin{eqnarray}
\label{crit-perc}
V_{\rm s}^{\rm c}
=
-
\frac
{g_2 n_2}
{\Omega_{\rm c} + \ln (1-\Omega_{\rm c})}
\;.
\end{eqnarray}
A numerical study for two-dimensional systems gives $\Omega_{\rm c} \approx 0.407$~\cite{PGMP10,PRBBAPAPS11}. This result is shown in Fig.~\ref{pd} by the dashed line and predicts qualitatively correct behavior of the phase boundary.

{\emph{Conclusions.---}}
Using mean-field theory and QMC calculations we have studied the superfluid fraction $\nu_{\rm s}$ as well as the fraction of particles with vanishing momentum $\nu_0$ for the interacting Bose gas at zero temperature in the presence of correlated disorder created by laser speckles in a quasi-2D setup. It is shown that $\nu_{\rm s}$ and $\nu_0$ smoothly decrease with the disorder strength $V_{\rm s}$ and there is a transition from the superfluid to the insulating regime. With the increase of interactions $a_{\rm s}$, $\nu_{\rm s}$ always grows while $\nu_0$ has in general a non-monotonic dependence. For large $a_{\rm s}$ and $V_{\rm s}$, $\nu_0$ can exceed $\nu_{\rm s}$. The results of the mean-field theory are in excellent agreement with the exact QMC calculations in the dilute regime even in the case of strong disorder. The predicted phase diagram can be accessed in the present-day experiments.

\emph{Acknowledgment.---}
We thank C.~A.~M\"uller for critical reading of the manuscript and helpful comments.
G.E.A. acknowledges fellowship by MEC (Spain) and financial support by (Spain) Grant No.~\abbrev{fis}2008-04403 and Generalitat de Catalunya Grant  No.~2009\abbrev{sgr}-1003. The work of K.V.K. and P.N. was supported by the SFB/TR 12 of the German Research Foundation (DFG). Part of the numerical calculations was carried out at the Supercomputing Centres in J\"ulich and Barcelona.


\end{document}